\documentclass[a4paper]{jpconf}
\usepackage{graphicx}

\begin{document}

\title{BCVEGPY and GENXICC for the hadronic production of the doubly heavy mesons and baryons}

\author{Xing-Gang Wu}

\address{Department of Physics, Chongqing University, Chongqing 401331, P.R. China}

\ead{wuxg@cqu.edu.cn}

\begin{abstract}
Doubly heavy mesons and baryons provide a good platform for testing pQCD. Two high efficient generators BCVEGPY and GENXICC for simulating the hadronic production of the doubly heavy mesons and baryons have been developed in recent years. In this talk, we present their main idea and their recent progresses. The dominant gluon-gluon fusion mechanism programmed in those two generators are written based on the improved helicity amplitude approach, in which the hard scattering amplitude are dealt with directly at the amplitude level and the numerical efficiency are greatly improved by properly decomposing the Feynman diagrams and by fully applying the symmetries among them. Moreover, in comparison to the previous versions, we have updated the programs in order to generate the unweighted meson or baryon events much more effectively within various simulation environments. The generators can be conveniently imported into PYTHIA to do further hadronization and decay simulation.
\end{abstract}

\section{Introduction}

Heavy quarkonium or baryon has attracted wide attention due to its special features, which provides a good platform for studying the perturbative QCD and the associated non-perturbative physics in the bound state system [1-4]. For example, the $B_c$ meson and the $\Xi_{cc}$ baryon can be a fruitful laboratory for testing various potential models and understanding the weak decay mechanism for heavy flavors. Systematic studies for the hadronic production of the doubly heavy quarkonium or baryon at the hadronic colliders as TEVATRON and LHC have been done in the literature [5-17]. In particular, two generators BCVEGPY and GENXICC have been completed and developed in recent years [18-24], which can be conveniently implemented into PYTHIA [25] for simulating the $B_c$ meson and $\Xi_{cc}$ baryon events with high efficiency. In the present paper, we present their main idea and their recent progresses. It is noted that in the case of productions of double heavy baryons the diquark is remaining in the coloured state. In this case the interactions with hadronic residue part may decrease the results of calculations due to disassociation effect [26]. Therefore, it is worth to mention, that the results of calculations of cross-sections of hadronic production of double heavy baryons are not absolutely identical to the case of production of $B_c$ meson and the obtained results for the calculations of the cross-sections of production of baryons should be considered with care and may be treated as the estimate of upper bound of the values of the cross-sections.

There are gluon-gluon fusion mechanism, or the light quark-anti-quark mechanism, or the extrinsic or intrinsic heavy quark mechanism for the hadronic production of the doubly heavy meson or baryon respectively. All of which may provide sizable contribution in specific kinematic regions. It is noted that in most of the kinematic regions (e.g. large $p_t$ region), the dominant production mechanism is the gluon-gluon fusion mechanism, e.g. for the production of the $(c\bar{b})$-quarkonium, it stands for the production via the process $gg\rightarrow (c\bar{b})[n]+b+\bar{c}$, where $[n]$ stands for the Fock-states $|(c\bar{b})_{\bf 1}[^1S_0]\rangle$, $|(c\bar{b})_{\bf 8}[^1S_0]g\rangle$, $|(c\bar{b})_{\bf 1}[^3S_1]\rangle$, $|(c\bar{b})_{\bf 8}[^3S_1]g\rangle$, $|(c\bar{b})_{\bf 1}[^1P_1]\rangle$ and $|(c\bar{b})_{\bf 1}[^3P_J]\rangle$ (with $J=(1,2,3)$) within the nonrelativistic QCD [27], respectively. The lowest Fock state $|(c\bar{b})_{\bf 1}[^1S_0]\rangle$ corresponds to the usually called $B_c$ meson. The cases for the hadronic production $\Xi_{cc}$, $\Xi_{bc}$ and $\Xi_{bb}$ baryons are similar, which can be obtained from the case of $B_c$ meson production by properly transforming the anti-quark line to be the quark line and by properly dealing with the color flows [13]. So, in the following, we shall take the gluon-gluon fusion mechanism for the $B_c$ meson production as an explicit example to explain our calculation technology and to show how to improve the events simulation efficiency.

\section{Gluon-gluon fusion mechanism for the hadronic production}

According to the pQCD factorization, the hadronic production cross section is formulated as
\begin{equation}
d\sigma=\int dx_{1}\int dx_{2} \; F^{g}_{H_{1}}(x_{1},\mu_{F})\times
F^{g}_{H_{2}}(x_{2},\mu_{F}) \; d\hat{\sigma}_{gg\rightarrow
(c\bar{b})[n] X}(x_{1},x_{2},\mu_{F},\mu_{R}), \label{pqcdf}
\end{equation}
where $F^{g}_{H}(x,\mu_{F})$ is the gluon distribution function in hadron $H$, $d\hat{\sigma}_{gg \rightarrow B_{c}X}(x_{1},x_{2},\mu_{F})$ is the cross section for the relevant inclusive production ($g+g\rightarrow (c\bar{b})[n]+b+\bar{c}$). $\mu_F$ and $\mu_R$ stand for the factorization scale and renormalization scale respectively. The generating of phase space and its integration can be done with the help of the routines RAMBOS [28] and VEGAS [29]. Then, the most important part left is to deal with the hard scattering amplitude of the process.

At the lowest $\alpha_s^4$ order, there are totally 36 Feynman diagrams and hence 36 amplitudes for the gluon-gluon fusion $g+g\rightarrow (c\bar{b})[n]+b+\bar{c}$. According to the pQCD factorization, each amplitude can be factorized into two parts, that of perturbative $gg \rightarrow b+\bar{b} +c+\bar{c}$ (all the quarks are on shell) and that of the nonperturbative $c+\bar{b} \rightarrow (c\bar{b})[n]$ which can be represented by the universal matrix element for each Fock state. Using the conventional squared amplitude to deal with the amplitude is a tedious and time-consuming task. To save the time for events simulation, we can deal with the hard scattering amplitude for $gg\rightarrow b+\bar{b}+c+\bar{c}$ directly at the amplitude level. In the literature, two ways have been suggested, i.e. the improved trace technology [30-33] and the helicity amplitude approach [34,35]. In our generators, we adopt an improved version for the helicity amplitude approach as suggested in Ref.[18] to simplify the hard scattering amplitude.

\subsection{A short review of the improved helicity amplitude approach}

\begin{figure}[h]
\includegraphics[width=20pc]{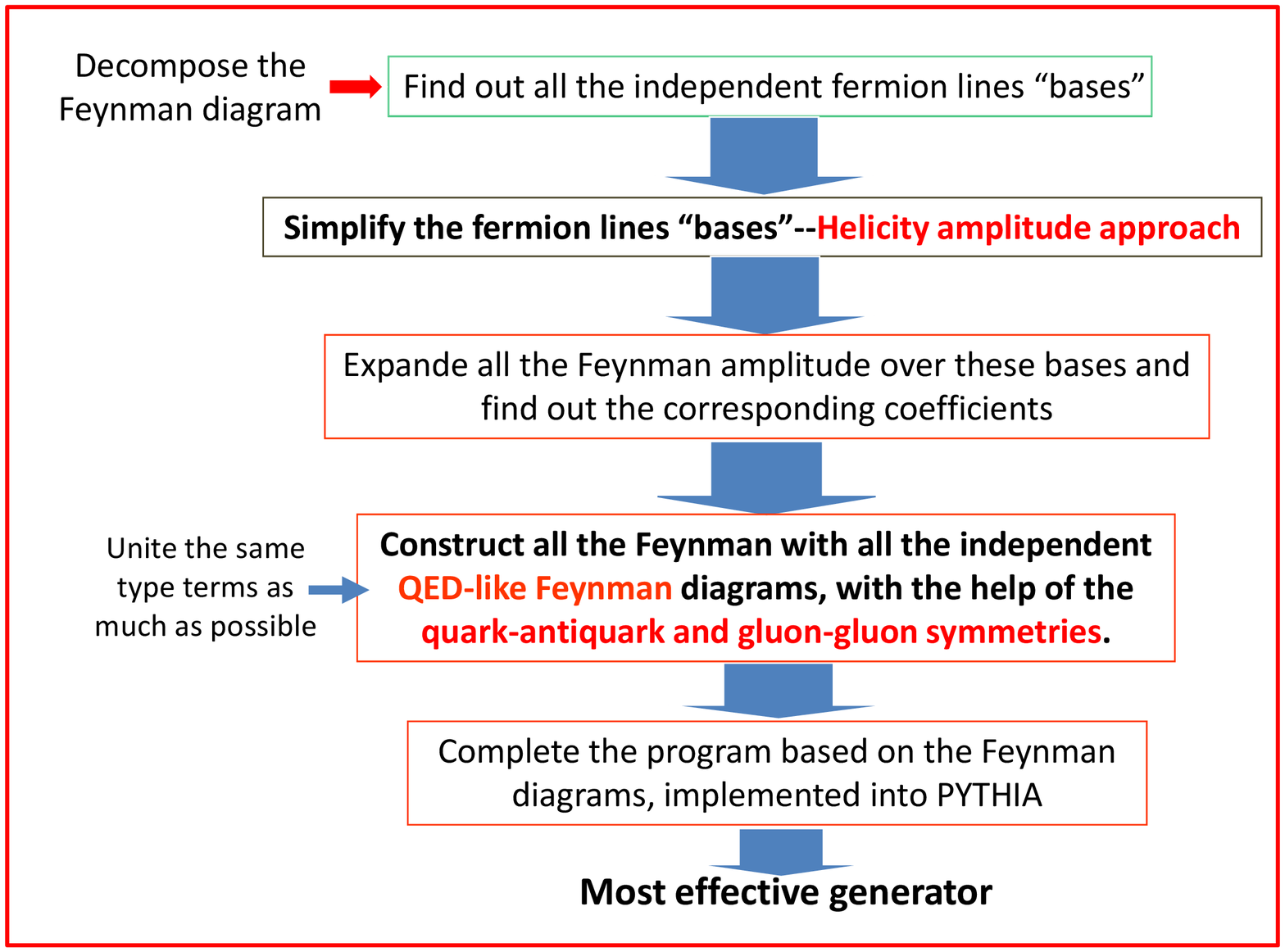}\hspace{2pc}%
\begin{minipage}[b]{15pc}
\caption{Main idea for the improved helicity amplitude approach [15].}
\end{minipage}
\end{figure}

Main idea of the improved helicity amplitude approach is shown in Fig.(1). According to the improved helicity amplitude approach [15], we first pick out all the basic spinor lines, which are constructed by the $\gamma$ structure of Fermion line, from the whole amplitude, whose remaining parts include the color factors and the scalar part of the propagators that shall be dealt with separately. These basic spinor lines can be simplified by changing the involved space-like momenta to be light-like ones, and then use the relation \(\slash\!\!\!k= |k+\rangle\langle k+|+|k-\rangle\langle k-|\) ($k$ is light-like) to transform and simplify all the spinor line into expressions of spinor products. This intermediate step can be done using the standard helicity amplitude approach, cf.Refs.[34,35]. Next, we construct the basic Feynman diagrams (nine for the gluon-gluon fusion) by proper combination of these basic spinor lines. Finally, we adopt the symmetries of the gluon-gluon exchanges, the quark line exchanges to construct all the amplitude of the process.

One subtle point. The authors of Ref.[34] group the Feynman diagrams of the process into gauge-invariant subsets. To achieve the goal, one have to introduce extra terms into the Feynman rules for the three-gluon and four-gluon vertices, i.e.
\begin{eqnarray}
gf_{abc}\tilde{T}^{\mu\nu\delta}(P,S,K)&=&gf_{abc}T^{\mu\nu\delta}(P,S,K)
+gf_{abc}G^{\mu\nu\delta}(P,S,K), \nonumber\\
-ig\tilde{V}_{abcd}^{\lambda\mu\nu\delta}(P,Q,K_{1},k_{2})&=&
- ig V_{abcd}^{\lambda\mu\nu\delta}(P,Q,K_{1},k_{2})-ig G_{abcd}^{\lambda\mu\nu\delta} (P,Q,K_{1},k_{2}),
\end{eqnarray}
where $P$, $S$ and etc. are input momenta for the gluons, $T^{\mu\nu\delta}(P,S,K)$ and  $V_{abcd}^{\lambda\mu\nu\delta}(P,Q,K_{1},k_{2})$ are primary Feynman rules
\begin{eqnarray}
T^{\mu\nu\delta}(P,S,K)&=& (P-S)^{\delta}g^{\mu\nu}+(S-K)^{\mu}g^{\delta\nu}
+(K-P)^{\nu}g^{\delta\mu}  \nonumber\\
V_{abcd}^{\lambda\mu\nu\delta}(P,Q,K_{1},k_{2})&=&
f_{abe}f_{cde}(g^{\lambda\nu}g^{\mu\delta}-g^{\lambda\delta}g^{\mu\nu})+
f_{ace}f_{dbe}(g^{\lambda\delta}g^{\mu\nu}-g^{\lambda\mu}g^{\nu\delta})\nonumber\\
&& + f_{ade}f_{bce}(g^{\lambda\mu}g^{\nu\delta}-g^{\lambda\nu}g^{\mu\delta}),
\end{eqnarray}
and $G^{\mu\nu\delta}(P,S,K)$ and $G_{abcd}^{\lambda\mu\nu\delta}(P,Q,K_{1},k_{2})$ are modified parts
\begin{eqnarray}
G^{\mu\nu\delta}(P,S,K)&=&(\pm)\left[P^{\mu}P^{\nu}P^{\delta}/(P\cdot K)+
S^{\mu}S^{\nu}S^{\delta}/(S\cdot K)\right] \\
G_{abcd}^{\lambda\mu\nu\delta}(P,Q,K_{1},k_{2})&=&
-f_{ace}f_{bde}\frac{S_{1}^{\lambda}S_{1}^{\mu}S_{1}^{\nu}S_{1}^{\delta}}
{(S_{1}\cdot K_{1})(S_{1}\cdot K_{2})}-f_{ade}f_{bce}\frac{S_{2}^{\lambda}S_{2}^{\mu}
S_{2}^{\nu}S_{2}^{\delta}}{(S_{2}\cdot K_{1})(S_{2}\cdot K_{2})},
\end{eqnarray}
where $S_{1}=P+K_{1}$ and $S_{2}=S+K_{2}$. Because the modified part of the three gluon vertex is symmetric on the interchange of $P$ and $S$, we must be careful to choose the sign of modified part in every associated diagrams so as to make all subsets be gauge invariant. When adding all extra terms together they will result in the required zero contribution. Then by choosing different gauge for each subset, one can obtain a compact result, when all terms of the amplitude are written according to the helicities of the fermions.

This method is most effective for massless cases. While, for the present massive case (the quark lines are massive), the question is much more involved. It is noted that the sum of all extra terms involving $G^{\mu\nu\delta}(P,S,K)$ and $G_{abcd}^{\lambda\mu\nu\delta}(P,Q,K_{1},k_{2})$ will have non-zero contributions, which complicates the final results. So in the improved helicity amplitude approach, we adopt a unique gauge for the whole amplitude. This gauge choice also avoids numerical cancelations between very large numbers and the numerical results are much more steady when doing the numerical calculations. This gauge independence can be adopted as one critical point on checking the rightness of the generator.

\begin{figure}[h]
\includegraphics[width=20pc]{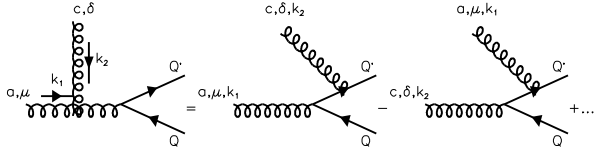}\hspace{2pc}%
\begin{minipage}[b]{15pc}
\caption{One typical three-gluon coupling vertex to be decomposed: the first two terms are the `basic QED-like' terms and the `remaining' terms are expressed by
several extra basic functions from the diagrams involving the four-gluon vertex.}
\end{minipage}
\end{figure}

Another subtle point. We can further make the amplitude more symmetric by applying a proper decomposition of the amplitude. As shown by Fig.(2), this is achieved by decomposing the terms involving three- and/or four-gluon vertices, into terms without self-interactions of gluons (i.e. into QED-like amplitudes). More over, it is noted that for the massive quark case, in addition to the QED-like Feynman diagrams, some more additional QED-like terms need to be introduced.

Fortunately, for the present subprocess $gg \rightarrow b+\bar{b} +c+\bar{c}$, these `extra' terms are just parts of the diagrams with four-gluon vertices; i.e. they are one of the three terms within the amplitude involving the four-gluon vertex. This shows that the whole amplitude is just the recombination of all the basic functions already existed in the 36 Feynman amplitudes, which is like a magic. There are totally nine basic QED-like Feynman diagrams for the present process. Thus, only the analytic expressions for those `basic Feynman diagrams' need to be put in the program precisely, while the non-basic ones can be generated by means of linear combination (the coefficients can be found in Ref.[18]) and/or by properly using the symmetry of all those basic ones.

\begin{figure}[h]
\includegraphics[width=22pc]{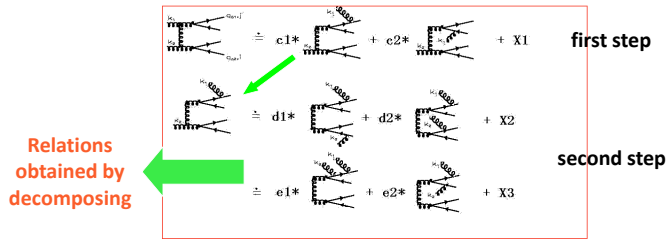}
\begin{minipage}[b]{15pc}
\caption{One example to decompose the Feynman diagram involving two three-gluon vertices to QED-like diagrams. Thus, we can obtain one relation among the QED-like diagrams.}
\end{minipage}
\end{figure}

As a final subtle point. As shown by Fig.(3), for the Feynman diagrams involving two three-gluon vertices, there are usually two ways to transform them into the QED-like diagrams; i.e. in the second step there are two ways to move the gluon either to the upper quark line or to the lower quark line respectively. Thus, two extra constraints can be found among the nine `basic Feynman diagrams', the program can be further simplified and become more compact.

\subsection{The structure and the efficiency of the BCVEGPY}

\begin{figure}[h]
\includegraphics[width=20pc]{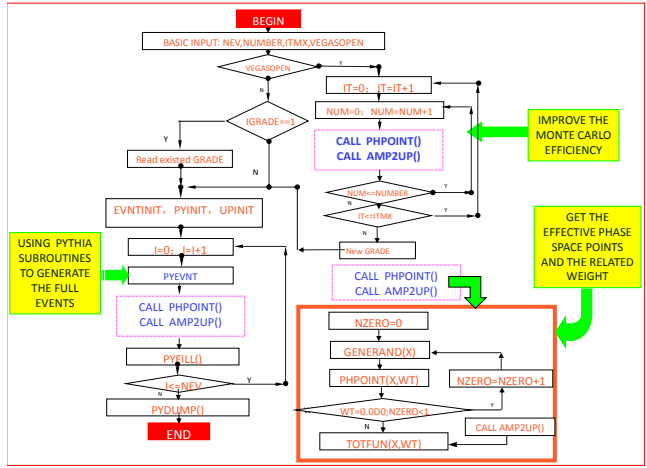}\hspace{2pc}%
\begin{minipage}[b]{15pc}
\caption{The flowchart for the generator BCVEGPY. The flowchart for the generator GENXICC is similar.}
\end{minipage}
\end{figure}

The flowchart for BCVEGPY, which is implemented into PYTHIA, is shown in Fig.(4). The BCVEGPY (or GENXICC) is a Fortran programme written in a PYTHIA-compatible format and is written in a modularization structure, one may apply it to various situations or experimental environments conveniently. It is noted that the LHE format is a standard format that is proposed to store process and event information from the matrix-element-based generators. One can pass these parton-level information to the event generator as PYTHIA for further simulation. And BCVEGPY will generate a standard LHE data file [36] that contains useful information of the meson and its accompanying partons, which can be conveniently imported into PYTHIA to do further hadronization and decay simulation.

\begin{figure}[h]
\includegraphics[width=20pc]{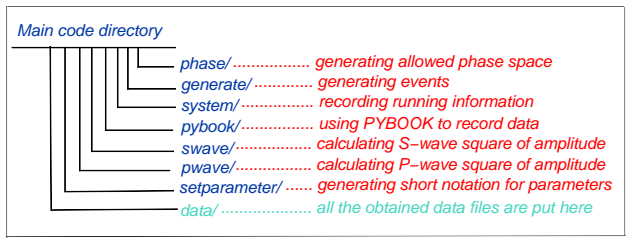}\hspace{2pc}%
\begin{minipage}[b]{15pc}
\caption{The schematic structure for the generator BCVEGPY. The schematic structure for the generator GENXICC is similar.}
\end{minipage}
\end{figure}

The schematic structure for the generator BCVEGPY is shown in Fig.(5). In general, the generator is systematically constructed in seven modules according to their purpose. Each module contains necessary files to fulfill the specific tasks for generating events. The BCVEGPY is dominated by two blocks, i.e. the {\it vegas} block (in module {\bf phase}) and the {\it event} block (in module {\bf generate}). The {\it vegas} block is to generate the importance sampling function. The {\it event} block is to generate events by using PYTHIA, in a way that all the mentioned processes are implemented into PYTHIA as its external processes. This is achieved by properly setting the two PYTHIA subroutines UPINIT and UPEVNT.

Here we modify the {\it vegas} block not only to generate the sampling importance function but also to store an upper bound of the value of the cross sections in each cell [37-40]; i.e. an improved and more effective method to generate unweighted events has been programmed [21,24] based on the MINT package [40]. The importance sampling function is used to increase the simulation efficiency, while the upper bound value will be used to generate unweighted event if the user want to do the experimental analysis and further simulation. The upper bound value in each cell is an upper bound for the cross sections and also equals a multidimensional stepwise function, according to which it is easy to generate phase-space points. By using this new hit-and-miss technique, one can generate the points according to the original distribution. Moreover, in doing the initialization, we will call VEGAS twice to generate the upper bound grid XMAX and also a more precise importance sampling function. In VEGAS the integral together with its numerical error are related to the sampling numbers {\rm ncall} and the iteration times {\rm itmx}. So, to generate full events, we suggest the user to do a test running first so as to find an effective and time-saving parameters for VEGAS. Finally, one can generate events by calling the UPEVNT subroutine according to the probability proportional to the integrand.

\section{Summary}

The coming LHC experiment shall provide a better platform to check all the theoretical predications and to learn $B_c$, $\Xi_{cc}$, $\Xi_{bc}$ and $\Xi_{bb}$ properties in more detail. Recently some new measurements on $B_c$ meson have been done by LHCb and CDF collaborations [41-45], in which BCVEGPY has been adopted for data analysis. Due to their high running efficiency (because of the using of improved helicity amplitude approach), BCVEGPY and GENXICC are very useful for Monte Carlo simulation and also for theoretical studies. At the present, they have been adopted by most of the groups at LHC and TEVATRON. The suggesting future super Z factory, GIGAZ program, LEP3, and etc., shall also provide good platforms for doubly heavy meson and baryon productions, cf. Refs.[46-49].

\section{Acknowledgments}

This talk was based on the works done in collaborations with Profs.Chang CH, Wang JX, Qiao CF and Dr.Wang XY. This work was supported in part by the Fundamental Research Funds for the Central Universities under Grant No.CQDXWL-2012-Z002, by Natural Science Foundation of China under Grant No.11075225 and No.11275280, and by the Program for New Century Excellent Talents in University under Grant No.NCET-10-0882.

\end{document}